\documentclass[11pt,twoside]{article}

\usepackage{asp2004}
\usepackage{epsf}
\usepackage{psfig}
\usepackage{lscape}

\begin{document}
\title{On the metallicity dependence of the winds from red supergiants and
       Asymptotic Giant Branch stars}          
\author{Jacco Th. van Loon}                    
\affil{School of Physical \& Geographical Sciences, Lennard-Jones
       Laboratories Keele University, Staffordshire ST5 5BG, United Kingdom}

\begin{abstract} 
Over much of the initial mass function, stars are destined to become luminous
and cool red giants. They may then be able to produce dust in an atmosphere
which has been elevated by strong radial pulsations, and hence drive a wind.
The amount of mass that is lost in this way can be a very significant fraction
of the stellar mass, and especially in the case of intermediate-mass stars it
is highly enriched. The delay between a star's birth and its feedback into the
environment varies from several million years for massive stars to almost the
age of the Universe for the least massive red giants we see today. I here
present a review on the metallicity dependence of red giant winds. I show that
recent measurements not only confirm theoretical expectations, but also
admonish of common misconceptions with implications for feedback at low
initial metallicity.
\end{abstract}



\section{An introduction to red giants}

Before discussing their stellar winds, I will first briefly introduce the red
giants themselves. Metalliciy is an important factor in determining the
structure and photospheric properties of these stars. Metal-poor stars are
generally more compact and have warmer photospheres because they lack opacity
in their mantles. But chemical enrichment of their mantles and photospheres
can influence that.

\subsection{Asymptotic Giant Branch stars with carbon+oxygen cores}

A star with an initial, zero-age main sequence mass $M_{\rm ZAMS}\sim1$ to 8
M$_\odot$ will at some point in its evolution switch from core-hydrogen
burning to hydrogen-shell burning, and then via a phase of core-helium burning
to hydrogen-shell burning again. During that final phase it will become
increasingly luminous as the energy-producing shell deposits its waste onto
the growing core, and as a reaction its mantle will expand and its surface
cool: the star ascends the Asymptotic Giant Branch (AGB). With a luminosity of
up to $L_{\rm AGB-tip}\sim6\times10^4$ L$_\odot$, AGB stars are powerful
beacons that can be used to probe intermediate-age populations between
$\sim30$ Myr and 10 Gyr old --- i.e.\ over much of the Universe's history.
Especially at infrared wavelengths the contrast with the underlying
main-sequence population and more massive, hotter stars is favourable.

An AGB star is most easily recognised if it shows regular variability in
brightness with a period of the order of a year or more at an amplitude that
can be several magnitudes at optical wavelengths, when it is called a ``Mira
variable'' after its peculiar prototype $o$\,Ceti (Mira). This variability is
explained by radial pulsation of the photosphere as a result of an instability
in the balance between the gravitation and radiation pressures, which arises
around the hydrogen and/or helium recombination zone(s) where the opacity
changes abruptly: the $\kappa$ mechanism (Christy 1962). The pulsation is most
vigorous if excited in the fundamental mode, when the rate at which energy is
deposited into the expanding mantle and released as the mantle contracts again
is highest.

On the upper slopes of the AGB, the pressure between the core and the
hydrogen-burning shell steadily grows until suddenly this helium-rich
interface layer ignites. The helium-burning shell eventually extinguishes
itself and the star switches back to hydrogen-shell burning. This repeats
itself on timescales of $10^4$ to $10^5$ yr: thermal pulses (Iben \& Renzini
1983). The effects, which are larger for less massive mantles, are twofold:
(1) the luminosity and temperature at the surface undergo an excursion either
way, and (2) the convective zone which occupies the outer mantle dips into the
fresh produce of nucleosynthesis, enabling these products to travel to the
stellar surface where they modify the photospheric abundance pattern. The
latter is called ``$3^{\rm rd}$ dredge-up'' (because two earlier phases of
surface enrichment are possible). This leads to enrichment in carbon and
products from the slow neutron capture process such as the unstable element
technetium, of which $^{99}$Tc with a half-life of $2\times10^5$ yr is an
unambiguous indicator of a thermal-pulsing AGB star.

The most dramatic result is obtained when the surface becomes so enriched in
carbon that the carbon atoms outnumber those of oxygen: a carbon star. This is
especially important because the surfaces of AGB stars are cool enough,
$T_{\rm eff}<4,000$ K, to form a molecular atmosphere. Although molecular
hydrogen, H$_2$, is the most abundant molecule in the presence of dust grains,
it is also the simplest and symmetric molecule; as such it is very hard to
notice by its opacity or emissivity. Most of the molecular opacity is due to
molecules involving either a carbon or an oxygen atom, due to their abundance
and high reactivity. It is therefore not surprising that carbon-monoxide, CO,
is the most common molecule after H$_2$. It also means that the molecular
chemistry is dominated by which ever is left after the formation of CO: oxides
such as H$_2$O (water), TiO and SiO, or carbonaceous molecules such as C$_2$,
CN and C$_2$H$_2$ (acetylene). Stellar models and observations suggest that
carbon stars are only produced for stars more massive than $M_{\rm
ZAMS}\sim1.3$ to 1.5 M$_\odot$, and that AGB stars more massive than $M_{\rm
ZAMS}\sim4$ M$_\odot$ do not become carbon stars because the carbon is burnt
to nitrogen at the bottom of the convection zone: Hot Bottom Burning (Marigo,
Girardi \& Bressan 1999; van loon, Marshall \& Zijlstra 2005).

\subsection{Red Supergiants}

In the cores of massive stars, with $M_{\rm ZAMS}>8$ M$_\odot$, eventually
carbon burning takes place. This creates an oxygen+neon core. At the lower
mass end, the evolution of these stars closely resembles that of AGB stars,
hence these are called ``super-AGB'' stars (Gil-Pons et al.\ 2005). More
massive stars will not undergo thermal pulses but they will continue to burn
ever heavier elements after core-carbon burning is exhausted until the iron
barrier induces core collapse. During core-helium burning, such a massive star
may become cool enough to also develop a convective, pulsationally unstable
mantle wrapped in a molecular atmosphere: a red supergiant (RSG), with a
luminosity $L_{\rm RSG}\sim10^5$ L$_\odot$.

\subsection{Red Giant Branch stars with electron-degenerate helium cores}

The helium cores of low-mass stars, with $M_{\rm ZAMS}<1.5$ to 2 M$_\odot$,
are electron-degenerate. With their cores so compact, in reaction their
mantles become relatively extended, making them already very cool during the
first phase of hydrogen-shell burning. As first-ascent Red Giant Branch (RGB)
stars they reach luminosities up to $L_{\rm RGB-tip}<3\times10^3$ L$_\odot$,
until core-helium burning ignites which ends the electron degeneracy in the
core. On the upper reaches of the RGB much of the photosphere is located in
the molecular atmosphere, and the convective mantle may undergo radial
pulsation in an overtone mode. Hence, like red supergiants, RGB stars share
similarities with AGB stars.

\section{Mass loss from red giants and the pivotal r\^{o}le of dust formation}

\subsection{Some evidence for mass loss from red giants}

Once the source of the luminosity inside stars was identified and the pieces
of the stellar evolution puzzle started falling into place, it became clear
that many stars will eject a significant amount of matter into space when they
die: supernova remnants in which the discovery of pulsars confirmed the
formation of neutron stars due to core collapse in massive stars, and
planetary nebulae in which the discovery of white dwarfs established the link
with lower-mass progenitors. More subtly, the properties of low-mass,
core-helium burning stars in globular clusters indicate that they may have 
already lost of order 20 per cent of their mass on the RGB (Woolf 1964; Rood
1973). And too many intermediate-mass stars would continue to evolve until the
explode as a core-collapse supernova unless this evolution were truncated by
the premature loss of the mantle, depriving the core of the fuel it needs to
accumulate mass.

Early spectroscopic observations of cool giant stars revealed violet-displaced
absorption in the optical line profiles of strong electronic transitions of
singly-ionized and neutral atoms. The corresponding bulk motion in the
atmosphere was believed to trace a steady outflow of matter, much like the
solar wind. Mass-loss rates were computed from the line profiles, and found to
be significant compared to the rate of evolution: $\dot{M}\sim10^{-8}$ to
$10^{-6}$ M$_\odot$ yr$^{-1}$ (Deutsch 1956; Reimers 1975). However, the
material was not seen to move at a speed that by itself would be sufficient
for the material to escape from the gravitational pull of the star and without
a certain mechanism for driving an outflow at some distance above the stellar
surface there was no certainty that this would indeed happen. Although larger
velocities are sometimes seen in the emission profile of H$\alpha$ the
interpretation is much more complicated, especially as the presence of a
chromosphere in all but the coolest carbon and late-M type stars gives rise to
significant thermal broadening. The strong pulsation of the coolest stars
causes large time-variable bulk motions in the photosphere, in both out and
inward directions (Hinkle, Hall \& Ridgway 1982), further complicating the
modelling of the base of an otherwise invisible wind.

\begin{figure}[!t]
\plotfiddle{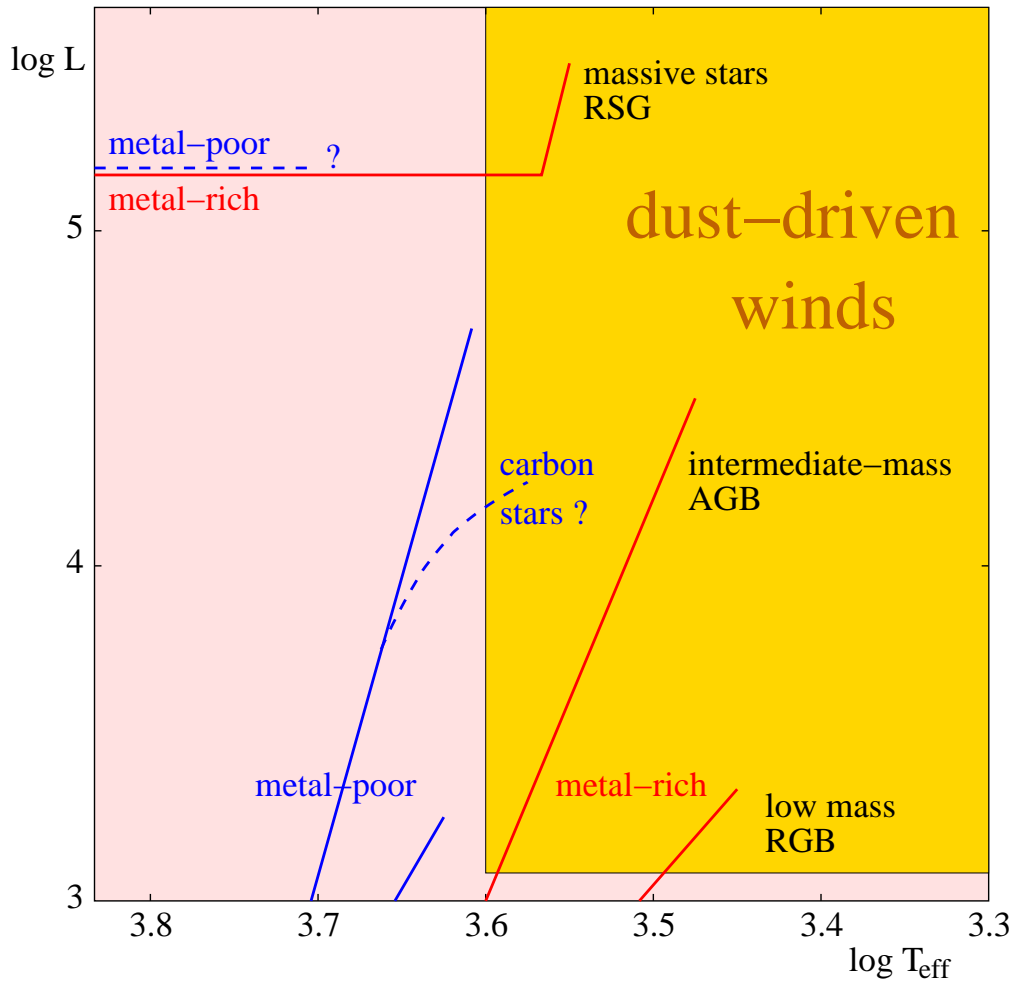}{130mm}{0}{58.5}{58.5}{-189}{-0}
\caption{Stars may evolve into a regime favourable to the development of
dust-driven winds, but this probably depends on mass and metallicity.}
\end{figure}

More direct evidence of a wind and mass loss from red giants came with the
detection of radio emission lines from abundant molecules, in particular the
amplified stimulated emission from hydroxyl (OH) at 1612 MHz. The
``classical'' double-peaked OH maser line profile, which resembles the Horns
of Consecration in Minoan Knossos, was soon explained by radial amplification
in a circumstellar envelope which expands at a constant speed (Elitzur,
Goldreich \& Scoville 1976). A dissociation product of water, OH is more
abundant downstream where the wind is more exposed to energetic interstellar
photons. The population inversion needed for the 1612 MHz satellite-line maser
is not possible in the inner, densest parts of the wind. These effects combine
to make the OH 1612 MHz maser an excellent probe of the steady outflow without
the complications near the base of the wind. Other maser lines include
transitions at 22 GHz in water vapour and at 43 and 86 GHz in SiO, both formed
closer to the star. The width of the thermal emission line profile of CO at
112 GHz was found to also measure the final wind speed, which enabled the
measurement of the wind speed for carbon stars as well as stars with
oxygen-rich envelopes (Knapp \& Morris 1985). An important observation is that
the wind does reach a velocity in excess of the local escape velocity, but not
until at least several stellar radii above the stellar photosphere (Richards
\& Yates 1998), implying a driving mechanism that is not limited to the
photospheric or sub-photospheric regions.

Advancement in technology gradually opened the infrared skies to astronomical
investigation, and it soon became clear that populations of stars exist which
are heavily reddened as a result of extinction at optical wavelengths by dust
grains. When observations at thermal infrared wavelengths ($\lambda>2$ $\mu$m)
became possible it was found that some of these stars have an excess of
infrared emission over that expected from a star behind a screen of
interstellar dust. This implies that the dust is very close to the star. The
association between these dust-enshrouded stars and circumstellar masers was
quickly made, and the dust emission became a very useful tracer of mass loss
(Gehrz \& Woolf 1971). The presence of dust also explained the pumping of the
masers through infrared emission, but most importantly it provided a mechanism
for driving a wind. The continuum opacity of the grains allows for an
efficient transfer of momentum from the stellar radiation field onto the dust
grains. As the photon momentum points away from the star it results in an
outward motion of the dust. If the density is high enough then the dust and
gas are collisionally coupled, mostly via grain-H$_2$ collisions. In this way,
the dust manages to carry with it more than hundred times its own weight in
gaseous matter. Very high mass-loss rates are estimated for these stars,
$\dot{M}\sim10^{-6}$ to $10^{-4}$ M$_\odot$ yr$^{-1}$ (Olnon et al.\ 1984).
The coincidence of cool photospheres and strong pulsation in these stars
inspired the theory that the pulsation acts as a piston to increase the
scaleheight of the molecular atmosphere, creating the temperature and density
regime required for the condensation of grains (Jura 1986).

\subsection{Measuring mass-loss rates via dust obscuration and re-emission}

Circumstellar dust grains absorb stellar light mainly at optical (and
ultraviolet) wavelengths, and re-emit it mainly at infrared wavelengths. The
shape of the observed spectral energy distribution (SED) depends on the
optical properties of the grains and on the optical depth of the envelope. For
a spherically symmetric geometry the integral under the SED yields the
bolometric luminosity, provided that the distance is known. The SED does {\it
not} allow a direct measurement to be made of the mass-loss rate. Under the
assumption of radiative equilibrium between the capture of photons and
isotropic emission by the heated grain, and applying the continuity equation,
one obtains a crude relationship that summarises the problem quite well
(Ivezi\'{c} \& Elitzur 1995):
\begin{equation}
\tau \propto \frac{\psi \dot{M}}{v_{\rm exp} \sqrt{L}},
\end{equation}
where we notice that although the optical depth, $\tau$, is proportional to
the (gas+dust) mass-loss rate, $\dot{M}$, it also depends on the dust:gas mass
ratio, $\psi$, the expansion velocity of the wind, $v_{\rm exp}$, as well as
the luminosity, $L$.

\section{Measurements of the metallicity dependence of red giant winds}

Mass loss of red giants has a big impact on their evolution, in particular on
the way they end their lives and the remnants they leave behind. The mass loss
has also a big impact on the evolution of their host galaxy. These stars
return a significant fraction of their mass, increasing the potential for
continued or invigorated star formation. The recycled material is chemically
enriched, supplying the interstellar medium (ISM) with metals that greatly
influence the chemical and thermal balance and the formation and evolution of
subsequent generations of stars. As grains cannot form under the conditions
encountered in the ISM, the circumstellar envelopes of red giants are of
particular importance in their r\^{o}le as dust factory. Because dust
condenses out of metals and seems to be an efficient way of driving mass loss
from red giants, and because of the interest in the early and metal-poor
Universe, the question arises how the winds from red giants depend on their
initial metallicity.

Measurement of the metallicity dependence is in principle possible within
components of the Milky Way. For instance in the galactic disk the scaleheight
of a stellar population increases with its age, and there is a general trend
for the metallicity to have increased over the assembly history of the
galactic disk. Thus, in general, stars further away from the mid-plane have a
lower metallicity than stars closer to the mid-plane (but they also have a
lower mass). It is difficult to measure accurate distances to RSGs, OH/IR
stars or carbon stars, many of which are either too far or optically too faint
for Hipparcos to have measured a parallax for. Although a relationship between
the pulsation period and luminosity exists for Mira-type AGB stars, stars may
deviate from the relationship, for instance due to mass loss lenghtening the
pulsation period. Hence it is difficult to measure the distance between a star
and the galactic mid-plane. This is true for any star, and measuring the
vertical metallicity gradient in the galactic disk is therefore a subject of
investigation of its own.

RGB stars can be studied in galactic globular clusters, which offer a range in
metallicity $0.01 < Z_{\rm globulars}/Z_\odot < 1$; but I shall concentrate
here on RSGs and AGB stars, in particular studies in the Magellanic Clouds
where the distances are known. Their initial metallicities in the Small (SMC)
and Large (LMC) Magellanic Clouds are $Z_{\rm SMC}\sim0.2$ Z$_\odot$ and
$Z_{\rm LMC}\sim0.4$ Z$_\odot$, making them ideal for studies of metallicity
dependence. Comparison with studies in the solar neighbourhood and the central
regions of our Galaxy extends the metallicity baseline to solar and
super-solar metallicity.

\subsection{Kinematics of dust-driven winds}

Apart from the fact that one needs to know the value for the wind speed in
order to derive the mass-loss rate from the SED (Eq.\ 1), a metallicity
dependence of the red giant wind speed may have implications for galactic
evolution. The wind speed is $v_{\rm exp}\sim5$ to 30 km s$^{-1}$ for AGB
stars and not much faster for RSGs. The wind momentum injection into the ISM
is therefore not very impressive, although it might contribute to the ISM
turbulence. Differences in wind speed may affect the timescale for mixing the
enriched material with the ambient ISM. With of order $10^2$ dusty AGB stars
per square kpc (Jura \& Kleinmann 1989), mixing timescales would be of order
$10^7$ yr --- short compared to the evolutionary and dynamical timescales of
these stars. But for the much rarer RSGs the mixing timescale is much longer
than the evolutionary timescales. The effect is greater in less flattened
systems with lower densities such as dwarf galaxies.

Simple radiation-driven dust wind theory predicts how the wind speed should
depend on the luminosity and dust:gas ratio. This not only allows the theory
to be tested but, if proven correct, it can also function as a scaling
relation to compute the expected wind speed for stars for which we have no
direct measurement of it. This then allows us to make a more realistic
estimate of the mass-loss rate from the optical depth than, as is often done,
by assuming a canonical value of, say, 10 or 15 km s$^{-1}$. The momentum
equation relates the motion of the mass and photon fluids:
\begin{equation}
\dot{M} v_{\rm exp} \propto \tau L,
\end{equation}
where the optical depth properly accounts for the scattering of photons off
the circumstellar grains. Hence, combination with Eq.\ (1) yields:
\begin{equation}
v_{\rm exp} \propto \sqrt{\psi} \sqrt[4]{L}.
\end{equation}
It seems deceptively reasonable that the dust:gas ratio depends on
metallicity. Unfortunately, this is a difficult parameter to measure directly.

First evidence for the metallicity dependence of the wind speed of red giants
was presented by Wood et al.\ (1992) who detected six OH/IR stars in the LMC,
of which four are AGB stars and the other two are RSGs. The wind speed for
these stars was found to be $\sim0.6$ of the values for similar stars in the
solar neighbourhood. Marshall et al.\ (2004) presented a larger sample of ten
OH/IR stars in the LMC, for eight of which values for the wind speed could be
estimated. They use a comparison with the wind speeds and luminosities of
OH/IR stars in the galactic centre to confirm Eq.\ (3) and indicate that the
dust:gas ratio is linearly proportional to metallicity:
\begin{equation}
\psi \propto Z.
\end{equation}
There is a suggestion in their data that the wind speed increases somewhat
more steeply with increasing luminosity, but it is unclear whether this is
real and if so, whether it is due to differences in the dust:gas ratio, or due
to an effect that is implicit in Eqs.\ (1) and (3).

Measurement of the wind speed in OH/IR stars in the SMC would provide a
sensitive test. There the winds are expected to contain very little dust,
which affects the efficiency of driving the wind, with a large drift velocity
between grains and gas. Eq.\ (3) may need to be modified to reflect this
(Habing, Tignon \& Tielens 1994). Detection of OH masers in SMC red giants is
an arduous affair; not only is the SMC a little more distant than the LMC but
also the reduced dust content in the circumstellar envelope lowers the mid-IR
flux that pumps the maser. Scaling the LMC population of OH/IR stars to the
size of the SMC further reduces the expected number of detectable OH masers in
the SMC.

In the galactic centre region, spheroidal and disk-like components meet (Wood
\& Bessell 1983). One way in which this population mixture manifests itself is
in the wind speed distribution of OH/IR stars: stars with $v_{\rm exp}>18$ km
s$^{-1}$ are identified with a metal-rich ($Z\sim3$ Z$_\odot$) population of
relatively young AGB stars, in addition to a bulge population of old,
metal-poor AGB stars with slower outflows (Wood, Habing \& McGregor 1998).

\subsection{Mass-loss rates of dust-driven winds}

The classical limit for the mass-loss rate of a radiatively-driven wind is
derived from the total conversion of photon momentum to mass momentum:
\begin{equation}
\dot{M}_{\rm classic}=\frac{L}{c v_{\rm exp}},
\end{equation}
where any excess absorbed photon energy is used to heat the grain. This does
not account for reflection of photons and multiple scattering, effects which
enhance the mass-loss rate. The general formula then becomes (Gail \& Sedlmayr
1986):
\begin{equation}
\dot{M}=\tau \dot{M}_{\rm classic},
\end{equation}
which can reach values much in excess of the classical limit. There has been
some confusion about the application of this result at different metallicity.
The optical depth depends on the dust:gas ratio but also on the wind speed
(Eq.\ 1), which itself depends on the dust:gas ratio and on which
$\dot{M}_{\rm classic}$ depends too. One could scale the dust:gas ratio with
metallicity how ever one wishes, yet it leaves the ratio $\tau/v_{\rm exp}$
and thus the mass-loss rate unchanged. {\it In a continuum-opacity driven wind
the mass-loss rate will not depend on metallicity}.

One can use Eqs.\ (1) and (2) to eliminate luminosity and obtain a combination
of dust:gas ratio and mass-loss rate as a function of optical depth. One can
obtain a similar, but different, combination when eliminating wind speed
instead of luminosity. This was used by van Loon (2000) to compare samples of
dust-enshrouded carbon stars and oxygen-rich AGB stars and RSGs in the SMC,
LMC, solar neighbourhood and galactic centre. For some of these stars the wind
speed was known, and for some of them the luminosity was known. The near-IR
colours were used to measure the optical depth. Internal consistency required
that the dust:gas ratio depends linearly on metallicity {\it and} that the
mass-loss rate depends at most very weakly on metallicity. This is true both
for the carbon stars and oxygen-rich red giants. Over an order of magnitude in
metallicity, $0.2 < Z < 3$ Z$_\odot$, there was no evidence for a significant
deviation from a metallicity independent mass-loss rate:
\begin{equation}
\frac{{\rm d} \dot{M}}{{\rm d} Z} = 0.
\end{equation}

\begin{figure}[!t]
\plotfiddle{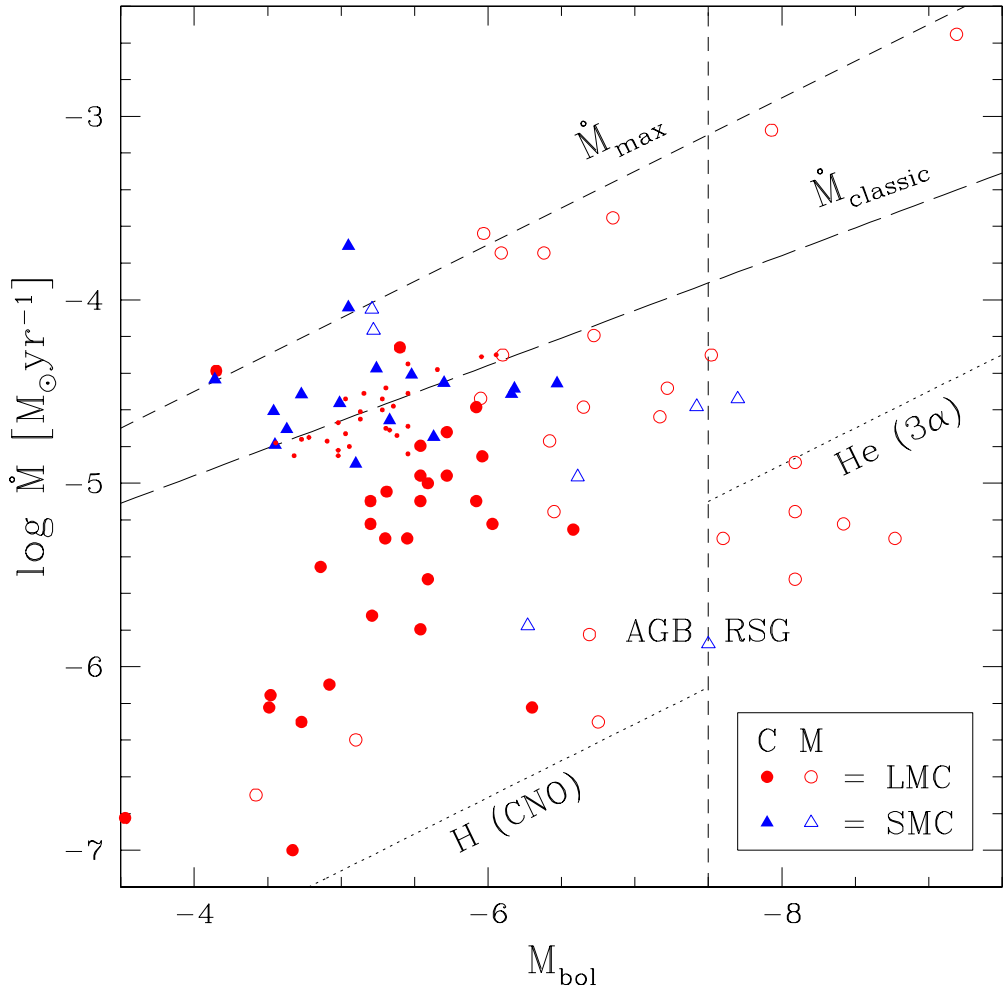}{130mm}{0}{69}{69}{-200}{-100}
\caption{Mass-loss rates versus bolometric magnitude for carbon stars and
oxygen-rich M-type AGB stars and RSGs in the LMC (circles: van Loon et al.\
1999; dots: van Loon et al.\ 2005b) and in the SMC (triangles: modified values
from Groenewegen et al.\ 2000).}
\end{figure}

The optical depth can be measured more reliably by modelling the SED with a
radiation transfer code, at near- and mid-IR wavelengths to consistently
reproduce the extinction and re-emission by the circumstellar dust envelope.
This was done by van Loon et al.\ (1999) for AGB stars and RSGs in the LMC to
draw a map of their distribution in mass-loss rate and bolometric luminosity.
A new sample of optically invisible carbon stars was selected and modelled by
van Loon et al.\ (2005b), mitigating previous bias by sampling low-luminosity
AGB stars with optically thick winds. The diagram (Fig.\ 2) shows a spread in
mass-loss rate due mainly to evolutionary effects, with the circumstellar
envelopes becoming optically thicker as the star cools and the pulsation
strengthens. The mass loss of AGB stars readily dominates stellar evolution
over the nuclear burning mass consumption rate, but for RSGs this is not
generally the case. Very significant mass loss of AGB stars and RSGs seems to
be possible during a brief episode at an extreme point in the evolution,
reaching maximum rates:
\begin{equation}
\dot{M}_{\rm max} \propto L.
\end{equation}

Mass-loss rates were derived in a similar manner for stars in the SMC by
Groenewegen et al.\ (2000). They did not, however, scale the dust:gas ratio
and wind speed, nor did they perform a comparison with the results obtained in
the LMC, so I will do that here. Motivated by the results described above, the
dust:gas ratio is scaled according to Eq.\ (4), $\psi_{\rm SMC}=0.001$, and
the wind speed is scaled according to Eq.\ (3). The oxygen-rich stars S\,10,
S\,15, S\,18 and S\,28 had been modelled with dust opacities from Volk \& Kwok
(1988), which I scale down by a factor of ten to bring them in line with more
normal opacities such as Draine \& Lee (1984). There is no distinction between
the mass-loss rates of dust-enshrouded carbon stars in the SMC and those in
the LMC (Fig.\ 2). The statistics for the luminous oxygen-rich stars are quite
poor, due in part to the steepness of the initial mass function, but again
there is no evidence for different mass-loss rates between the Magellanic
Clouds.

The degree of dust condensation seems to approach unity for red giants in the
extreme mass-loss phase: only the metal abundance limits the amount of dust
that forms. But this relies on a sufficiently high density in the dust
formation zone, and thus on the pulsation piston that elevates the molecular
atmosphere. Although most studies concentrate on the pulsation period, it is
the pulsation amplitude that quantifies the rate at which energy is being
stored and released in the expansion and contraction of the stellar
photosphere (van Loon et al.\ 2005b). Stars in the Magellanic Clouds, galactic
bulge and galactic centre all seem to run against the maximum possible
pulsation energy injection rate, $\dot{E}_{\rm puls}$: the photosphere cannot
store more energy than what it would normally have radiated away during that
time (van Loon 2002). The maximum $\dot{E}_{\rm puls} \propto L$. It thus
seems natural to conclude that, as the dust condensation process has already
saturated, the maximum mass-loss rate of red giants is set by saturation of
the pulsation mechanism and is found to be largely metallicity independent:
\begin{equation}
\frac{{\rm d} \dot{M}_{\rm max}}{{\rm d} Z} = 0.
\end{equation}

\subsection{Dust and molecule formation}

Differences in the dust opacity, due to composition and grain size and shape,
affect the dynamics of the wind but also the value for the mass-loss rate
inferred from the measured optical depth. Carbon-rich dust is mostly composed
of amorphous carbon and a small fraction of silicon-carbide, SiC, whilst
oxygen-rich dust is predominantly composed of amorphous silicates, SiO$_4$ or
SiO$_3$, with inclusions of magnesium and iron (Ivezi\'{c} \& Elitzur 1995).
The warmest dust is probably in the form of alumina, Al$_2$O$_3$. Significant
fractions of crystalline dust are sometimes found in post-AGB objects. In
carbon star winds the paradigm is to form grains by growing carbon chains,
where acetylene plays a determining r\^{o}le. In oxygen-rich environments one
could expect that SiO goes on to form silicate grains, but chemistry
considerations require a pre-existing condensation nucleus such as TiO
molecules or perhaps alumina (Salpeter 1974).

Groenewegen et al.\ (1995) obtained groundbased 8--13 $\mu$m spectra of one
oxygen-rich dust-enshrouded star in each of the Magellanic Clouds and compared
the silicate features with that of a galactic OH/IR star. The optical depth
was measured to be highest in the Galaxy and lowest in the SMC, but the
quality of these pioneering data were insufficient to make statements about
the properties of the grains. With ISO, more and better mid-IR spectra were
obtained of samples of dust-enshrouded red giants in the LMC (Trams et al.\
1999) and in the SMC (Groenewegen et al.\ 2000), but there was no evidence for
differences in the dust properties between the SMC, LMC and Milky Way other
than a diminishing optical depth at lower metallicity. Several Spitzer Space
Telescope programmes have obtained 5--35 $\mu$m spectra of magellanic carbon
stars and oxygen-rich massive AGB stars and RSGs. Sloan et al.\ (2006) present
the first results, for carbon stars in the SMC, suggesting a smaller fraction
of SiC in the dust as compared to what is seen in the Milky Way, probably due
to the lower silicon abundance. SiC is a minor fraction of the dust in any
case, usually less than 10 per cent, and not very important for driving the
mass loss.

Certain molecular bands in magellanic carbon stars are found to be very strong
compared to galactic carbon stars, which has been explained by higher C:O
ratios when the oxygen abundance is lower but the carbon production by the
star itself remains efficient (van Loon, Zijlstra \& Groenewegen 1999;
Matsuura et al.\ 2002, 2005; van Loon et al.\ 2005b). This leads to diminished
abundances of HCN and CS which depend on the availability of nitrogen and
sulphur, respectively, but enhanced abundances of acetylene because less
carbon is locked up in CO. It is not yet clear whether this would also lead to
a higher dust:gas ratio in metal-poor carbon stars. If it does, then the
mass-loss rate may not be any different, but the wind speed will be higher.

Silicate grains require oxygen and silicon in the ratio O:Si=3 to 4. The solar
abundance pattern (Anders \& Grevesse 1989) has O:Si=24, and O:Si=14 after
creating CO. One could conclude that dust production in oxygen-rich AGB stars
and RSGs is limited by the amount of silicon not oxygen. As silicon is not
produced in the star itself, the dust:gas ratio would then saturate at a value
$\psi \propto Z$ --- just what is seen in OH/IR stars in the LMC and the
Galaxy. However, the C:O ratio can be closer to unity than in the Sun, which
would reduce the O:Si ratio. In massive AGB stars oxygen, like carbon, is
depleted by burning into nitrogen during Hot Bottom Burning, which also
reduces the O:Si ratio. Finally, a significant fraction of oxygen is used in
forming molecules other than CO and SiO, such as water. Thus it is possible,
in principle, that O:Si$<$3 and the dust production is limited by oxygen. Yet
again, a lower dust:to gas ratio may leave the mass-loss rate unchanged but it
will lower the wind speed.

\subsection{What happens at very low metallicity?}

Bowen \& Willson (1991) and Zijlstra (2004) suggest that dust-driven mass loss
breaks down at $Z<0.1$ Z$_\odot$. There can be two reasons why a dust wind
might fail to develop. Firstly, the dust:gas ratio must be large enough for
the radiation pressure to balance gravity in the dust-formation region, or the
envelope collapses. Secondly, if the density is too low then the dust
decouples from the gas and may be blown out of the system leaving most of the
mass to fall back onto the star. It is not clear whether the latter would
really happen, as the dust can only form in a relatively dense environment in
the first place.

A very strong temperature dependence of the mass-loss rate is predicted
(Wachter et al.\ 2002) and observed (Alard et al.\ 2001). The empirical
formula derived from dust-enshrouded oxygen-rich AGB stars and RSGs in the LMC
(van Loon et al.\ 2005a) also correctly predicts the observed mass-loss rates
for similar stars in the solar neighbourhood, but it overestimates the
mass-loss rates for less extreme galactic stars such as Betelgeuse which is
relatively warm and known for its very low dust:gas ratio. This merely
reflects the transition region between the domain of the dust-driven winds
around the coolest fundamental-mode pulsators, and the more stable dust-free
stars where the density is too low or timescale too short for complete
condensation to occur. The absence of Mira variables at $Z<0.1$ Z$_\odot$
(Frogel \& Whitelock 1998) suggests that metal-poor red giants do not
compensate for their warmer photosphere by pulsating more vigorously. One
therefore expects that metal-poor stars form dust later in their evolution
than metal-rich stars would have done. Very metal-poor stars may never form
dust, except perhaps if they become carbon star (Fig.\ 1).

On the other hand, the warmer red giants exhibit a chromosphere which may,
somehow, drive the mass loss that was originally measured in the optical lines
and which is the basis for the famous Reimers' law. This mechanism may replace
the r\^{o}le of dust-driven winds for (very) metal-poor stars. According to
the model proposed recently by Schr\"{o}der \& Cuntz (2005), the
chromospherically-driven mass-loss rate increases for warmer red giants.
Although they do not point this out themselves, one would come to the
remarkable conclusion that {\it metal-poor stars loose more mass than
metal-rich stars}.

\section{Summary and outlook on the future}

Observations support the paradigm that as long as a luminous giant star is
able to develop a cool, molecular atmosphere it will also pulsate and develop
a dust-driven wind. The dust:gas ratio is then irrelevant for the mass-loss
rate, although it does determine the wind speed. Although metal-poor stars may
not reach that stage, the alternative mechanism of chromospherically-driven
mass loss may in fact be more efficient for them, and total mass-loss rates
may not be lower than for metal-rich red giants.

That all is not that simple is plainly demonstrated, for instance by the
surprising observation of dust in metal-poor globular clusters (Origlia et
al.\ 2002; Evans et al.\ 2003). Because the nearby Universe does not offer a
direct means of observing supergiants or massive AGB stars at $Z<0.01$
Z$_\odot$, it is essential that a complete {\it theory} for the mass loss from
red giants be established. Until then, there are plenty of ways in which {\it
observations} can further our understanding of the red giant winds. More
precise measurements of the optical depth in large samples of red giants in
the Magellanic Clouds and the nearest metal-poor dwarf spheroidal galaxies
might resolve differences in the {\it inferred} mass-loss rates (such as in
Fig.\ 2) between these galaxies for different scaling laws of the dust:gas
ratio. Future measurements of CO in the envelopes around magellanic red giants
by ALMA will provide values for the wind speed and dust:CO ratio in both
carbon stars and oxygen-rich AGB stars and RSGs. This will greatly improve the
accuracy of the derived mass-loss rates and test the dust formation and wind
acceleration mechanisms. Measurements of the carbon and oxygen abundances in
massive AGB stars as a function of metallicity are possible and needed to test
theories that predict them to become carbon stars at very low metallicity.
Chromospherically-driven winds need to be studied for metal-poor AGB stars and
RSGs to gauge whether this would be a viable alternative to dust-driven winds.
It is too early to discard the r\^{o}le of red giants in the Early Universe.

\acknowledgements 
I would like to thank the scientific and local organisers and all participants
for a very pleasant and stimulating meeting in a beautiful country, and for
inviting me to present this review. I am also indebted to the Royal Society
for the award of a conference grant, and to Joana for many reasons.


\end{document}